\def\year{2022}\relax
\documentclass[letterpaper]{article} 
\usepackage{aaai22}  
\usepackage{times}  
\usepackage{helvet}  
\usepackage{courier}  
\usepackage[hyphens]{url}  
\usepackage{graphicx} 
\usepackage{natbib}  
\usepackage{caption} 
\usepackage{subfigure}
\usepackage[utf8]{inputenc}
\usepackage[T1]{fontenc} 
\usepackage{hyperref} 
\usepackage{url}       
\usepackage{booktabs}
\usepackage{threeparttable}
\usepackage{amsfonts}   
\usepackage{nicefrac}   
\usepackage{microtype}      
\usepackage{xcolor}         
\usepackage{amsmath}
\usepackage{graphicx}
\usepackage{amsmath}
\DeclareMathOperator*{\argmax}{arg\,max}
\DeclareMathOperator*{\argmin}{arg\,min}
\newcommand{\Tref}[1]{Table~\ref{#1}}

\newcommand{\Fref}[1]{Fig.~\ref{#1}}

\def\abs{\text{abs}}

\def\ie{\emph{i.e.}}
\def\eg{\emph{e.g.}}

\def\eg{{\emph{e.g.}}}
\def\ie{{\emph{i.e.}}}

\newcommand{\renjie}[1]{{#1}}
\newcommand{\yufei}[1]{{#1}}
\newcommand{\wh}[1]{{#1}}

\DeclareCaptionStyle{ruled}{labelfont=normalfont,labelsep=colon,strut=off} 
\usepackage{algorithm}
\usepackage{algorithmic}

\usepackage{newfloat}
\usepackage{listings}
\floatstyle{ruled}
\newfloat{listing}{tb}{lst}{}
\floatname{listing}{Listing}
\title{Low-Light Image Enhancement with Normalizing Flow}
\author {
    Yufei Wang, \textsuperscript{\rm 1}
    Renjie Wan, \textsuperscript{\rm 1}
    Wenhan Yang, \textsuperscript{\rm 1}
    Haoliang Li, \textsuperscript{\rm 2}
    Lap-Pui Chau, \textsuperscript{\rm 1}
    Alex C. Kot, \textsuperscript{\rm 1}
}
\affiliations {
    \textsuperscript{\rm 1} Rapid-Rich Object Search Lab, Nanyang Technological University, Singapore\\
    \textsuperscript{\rm 2}  Department of Electrical Engineering, City University of Hong Kong, China\\
    \{yufei,rjwan, wenhan.yang,elpchau,eackot\}@ntu.edu.sg, haoliang.li@cityu.edu.hk
}

\usepackage{bibentry}

\begin{document}

\maketitle

\begin{abstract}
To enhance low-light images to normally-exposed ones is highly ill-posed, namely that the mapping relationship between them is one-to-many. Previous works based on the pixel-wise reconstruction losses and deterministic processes fail to capture the complex conditional distribution of normally exposed images, which results in improper brightness, residual noise, and artifacts. In this paper, we investigate to model this one-to-many relationship via a proposed normalizing flow model. 
An invertible network that takes the low-light images/features as the condition and learns to map the distribution of normally exposed images into a Gaussian distribution. In this way, the conditional distribution of the normally exposed images can be well modeled, and the enhancement process, \ie. the other inference direction of the invertible network, is equivalent to being constrained by a loss function that better describes the manifold structure of natural images during the training.
The experimental results on the existing benchmark datasets show our method achieves better quantitative and qualitative results, obtaining better-exposed illumination, less noise and artifact, and richer colors.
\end{abstract}
\section{Introduction}
\label{sec:introduction}

\begin{figure}[t]
    \scalebox{1}{
    \centering
    \includegraphics[width=1\linewidth, trim=6 11 6 6, clip]{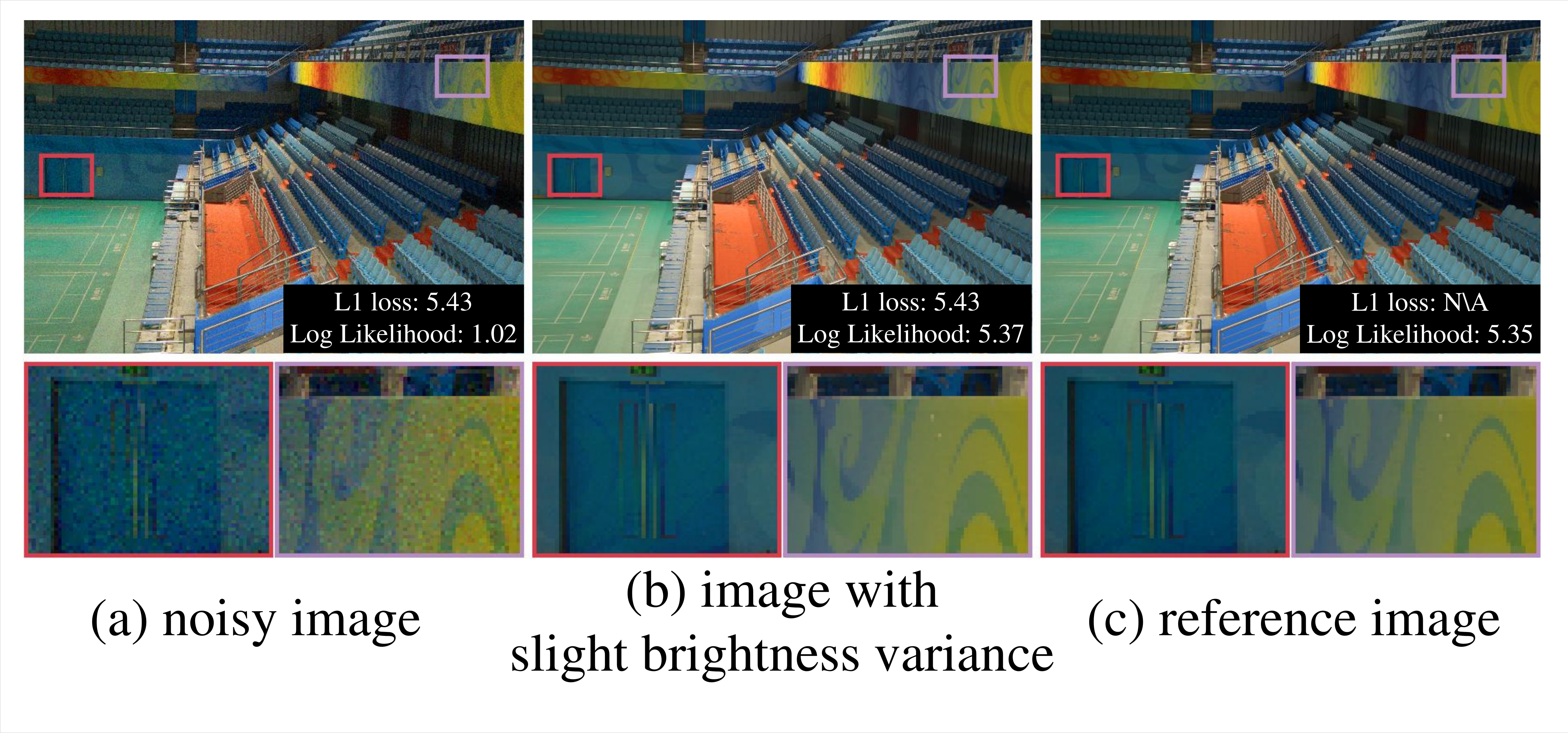}
    }
    \caption{
    \yufei{
    \wh{
    Illustration of the superiority of our normalizing flow model in measuring the visual distance compared to $l_1$ loss for low-light image enhancement.
    Although (b) is more visually similar to (c), \ie, reference image, than (a), their $l_1$ losses are the same.
    Benefiting from better capturing the complex conditional distribution of normally exposed images, our model can better capture the error distribution and therefore provide the measure results more consistent with human vision.
    }
    %
    %
    %
    }
    }
    \label{fig:cond-prob}
\end{figure}

Low-light image enhancement aims to improve the visibility of low-light images and suppress captured noise and artifacts.
Deep learning-based methods~\cite{zhang2019kindling, zamir2020learning,chen2018learning} achieve promising performance by utilizing the power of large collections of data.
However, most of them mainly rely on the pixel-wise loss functions (\eg, $l_1$ or $l_2$) in the network training that derive a deterministic mapping between the low-light and normally exposed images.
This enhancement paradigm encounters two issues.
First, this pixel-wise loss cannot provide effective regularization on the local structures in diverse contexts.
%
\yufei{As one low-light image may correspond to several reference images with different brightness \cite{zhang2021beyond}}, 
this pixel-to-pixel deterministic mapping is easily trapped into the ``regression to mean'' problem and obtains the results that are the fusion of several desirable ones, which inevitably leads to improperly exposed regions and artifacts.
Second, due to the simplified assumption of the pixel-wise losses about the image distribution, these losses might fail in describing the real visual distance between the reference image and enhanced images in the image manifold \yufei{as shown in~\Fref{fig:cond-prob}},
which further undermines the performance.
Though the \yufei{GAN-based} scheme can partly alleviate this issue, these approaches require careful tuning during training~\cite{wolf2021deflow} and might overfit certain visual features or the properties of the training data.

 
 
\yufei{Recently, researchers have shown the effectiveness of normalizing flow in the field of computational photography. \cite{wolf2021deflow, lugmayr2020srflow, xiao2020invertible}}
The normalizing flow is capable to learn a more complicated conditional distribution than the classical pixel-wise loss, which can well solve the above-mentioned two issues.
Beyond previous CNN-based models that learn a deterministic mapping from the low-light image to an image with specific brightness,
the normalizing flow learns to map the multi-modal image manifold into a latent distribution.
Then, the loss enforced on the latent space equivalently constructs an effective constraint on the enhanced image manifold.
It leads to better characterization of the structural details in various contexts and better measurement of the visual distance in terms of high-quality well-exposed images, which helps effectively adjust the illumination and suppress the image artifacts.
%
%
However, since the classical normalizing flow is biased towards learning image graphical properties such as local pixel correlations~\cite{kirichenko2020normalizing}, it may fail to model some global image properties like the color saturation, which can undermine the performance when applying these methods for the low-light image enhancement problem.

\begin{figure*}[t]
    \centering
    \includegraphics[width=1.\linewidth,trim=10 10 330 10 10,clip]{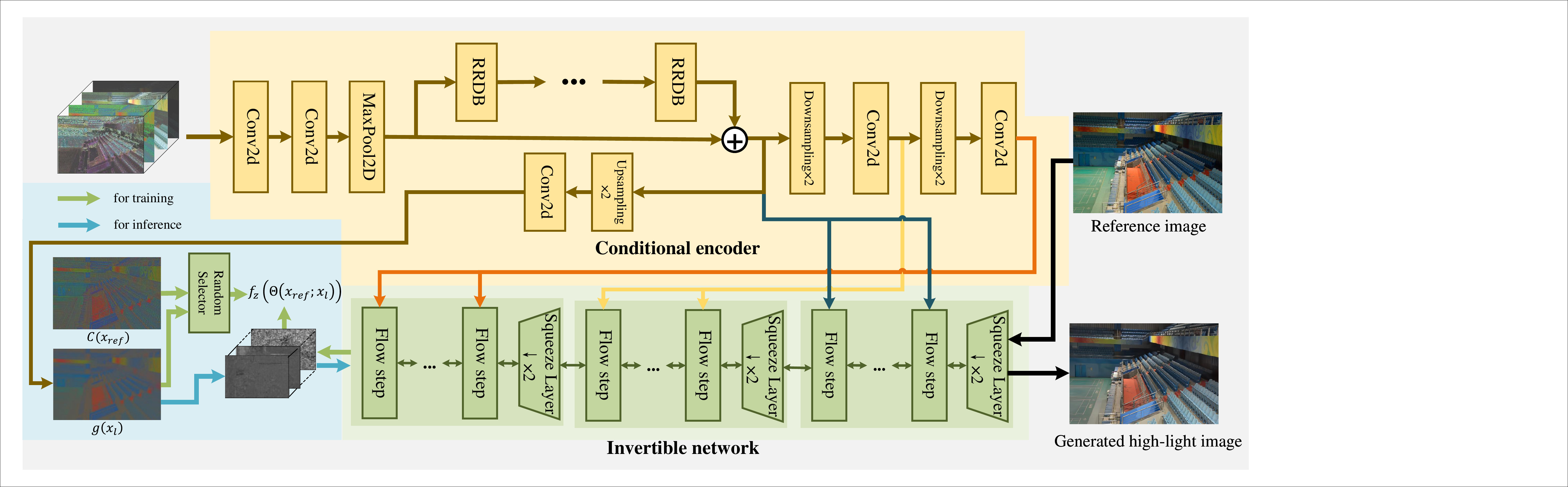}
    \caption{The architecture of our proposed LLFlow. Our model consists of a conditional encoder to extract the illumination-invariant color map and an invertible network that learns a distribution of normally exposed images conditioned on a low-light one. For training, we maximize the exact likelihood of a high-light image $x_h$ by using change of variable theorem in Eq. \wh{\eqref{eq:f_flow}} and a random selector is used to obtain the mean value of latent variable $z$  which obey Gaussian distribution \yufei{from the color map $C(x_h)$ of reference image or the extracted color map $g(x_l)$ from low-light image through the conditional encoder}. For inference, we can randomly select $z$ from $\mathcal{N}(g(x_l),\mathbf{1})$ to generate \yufei{different normally exposed images}
    from the learned conditional distribution $f_{flow}(x|x_l)$.
    (The color maps in the blue area are squeezed to the same size \wh{with} latent feature $z$.)}
    \label{fig:framework}
    \vspace{-0.3cm}
\end{figure*}

To address the above issues, in this paper, we propose \textbf{LLFlow}, a flow-based low-light image enhancement method to accurately learn the local pixel correlations and the global image properties by modeling the distributions over the normally exposed images.
As shown in~\Fref{fig:framework}, to merge the global image information into the latent space, instead of using standard Gaussian distribution as the prior of latent features, we propose to use the illumination-invariant color map as the mean value of the prior distribution.
More specifically, the encoder is designed to learn a one-to-one mapping to extract the color map that can be regarded as the intrinsic attributes of the scene that do not change with illumination. 
Simultaneously, another component of our framework, the invertible network, is designed to learn a one-to-many mapping from a low-light image to distribution of normally exposed images. As such, we expect to achieve better low-light image enhancement performance through our proposed framework.

In summary, contributions can be concluded as follows.
\begin{itemize}
    \item We propose a conditional normalizing flow to model the conditional distribution of normally exposed images. 
    It equivalently enforces an effective constraint on the enhanced image manifold. Via better characterization of the structural details and better measurement of the visual distance, it better adjusts illumination as well as suppresses noise and artifacts.
    %
    \item 
    We further introduce a novel module to extract the illumination invariant color map inspired by the Retinex theory as the prior for the low-light image enhancement task, 
    which enriches the saturation and reduces the color distortion.
    
    
    
    \item 
    We conduct extensively experiments on the popular benchmark datasets to show the effectiveness of our proposed framework.
    The ablation study and related analysis show the rationality of each module in our method.
\end{itemize}



\section{Related works}
\subsection{Low-light image enhancement}
As an active research topic in the past several years, a large number of low-light image enhancement methods have been proposed. Early methods mainly utilize the Retinex theory to correct the image illumination and suppress the artifacts. Recently, with the emergence of deep learning schemes, more tasks have benefited from the deep learning model. For example,  LLNet \cite{lore2017llnet} uses a deep auto-encoder to adaptively enlighten the image. Multi-scale features are adopted \cite{shen2017msr, tao2017llcnn, lv2018mbllen, ren2019low} to obtain better visual quality. The \cite{shen2017msr} illustrates the close relationship between Retinex and CNN with Gaussian convolution kernels, two separated deep networks are used for decomposition in \cite{wei2018deep}, and \cite{wang2019progressive} propose a progressive Retinex framework that the illumination and reflection maps are trained in a mutually reinforced manner. 
In addition, different losses are used to guide the training, e.g.,  MSE \cite{lore2017llnet, cai2018learning}, $l_1$ loss\cite{cai2018learning}, structural similarity (SSIM) \cite{cai2018learning}, smoothness loss \cite{wang2019underexposed,zhang2019kindling} and color loss \cite{wang2019underexposed, guo2020zero, shen2017msr}. Meanwhile, \cite{cai2018learning} demonstrates that training the same network with different reconstruction losses will have different performances which demonstrates the significance of conditional distribution design. Introducing carefully designed color loss can be also regarded as refining the conditional distribution, 
\ie, give color distortion pictures a greater penalty coefficient. 
Different from previous works that carefully design the reconstruction loss for end-to-end training, in this paper, we propose to utilize a normalizing flow to build the complex posterior distribution which has proven to be more effective and can generate images with higher quality, less noise, and artifact.  

\subsection{Normalizing flow}
A normalizing flow is a transformation of a simple probability distribution (\eg, a standard normal) into a more complex distribution by a sequence of invertible and differentiable mappings \cite{kobyzev2020normalizing}. Meanwhile, the probability density function (PDF) value of a sample can be exactly obtained by transforming it back to the simple distribution. To make the network invertible and computation tractable, the layers of the network need to be carefully designed so that the inversion and the determinant of Jacbian matrix can be easily obtained which limits the capacity of the generative model. To this end, many powerful transformations have been proposed to enhance expressiveness capacity of the model. For example, affine coupling layers~\cite{dinh2014nice}, split and concatenation~\cite{dinh2014nice, dinh2016density, kingma2018glow}, Permutation~\cite{dinh2014nice, dinh2016density, kingma2018glow}, and 1 $\times$ 1 convolution~\cite{kingma2018glow}. Recently, conditional normalizing flows are investigated to improve the expressiveness of the model. \cite{trippe2018conditional} propose to use different normalizing flows for each condition. Recently, conditional affine coupling layer~\cite{ardizzone2019guided,winkler2019learning,lugmayr2020srflow} is used to build a stronger connection with the conditional feature and improve the efficiency of memory and computational resource. Benefiting from the development of normalizing flow, the scope of application has been greatly expanded. For instance, \cite{liu2019conditional} generates faces with specific attributes, \cite{pumarola2020c, yang2019pointflow} use conditional flow to generate point clouds. In the super-resolution tasks, \cite{lugmayr2020srflow, winkler2019learning, wolf2021deflow} generate the distribution of high-resolution images based on one low-resolution input based on the conditional normalizing flow. Besides, the conditional normalizing flow is also used in image denoising \cite{abdelhamed2019noise,liu2021invertible} to generate extra data or restore the clean image.  
In addition, the inductive biases of normalizing flows are explored \cite{jaini2020tails, kirichenko2020normalizing}. \cite{kirichenko2020normalizing} reveals that the normalizing flow prefers to encode simple graphical structures which may be helpful to suppress the noise in the low-light image.

\section{Methodology}
In this section, we first introduce the limitations of previous \wh{pixel-wise} reconstruction loss-based low-light enhancement methods.
\wh{
Then, the overall paradigm of our framework in Fig. \ref{fig:framework} is introduced.
Finally, two components of our proposed framework are illustrated separately.}

\subsection{Preliminary}
The goal of low-light image enhancement is to generate a high-quality image with normal exposure $x_h$ using a low-light image $x_l$. Paired samples $(x_l, x_{ref})$ are usually collected to train a model $\Theta$ {by minimizing} the $l_1$ reconstruction loss as follows:
\begin{equation}
    \argmin_{\Theta} \mathbb{E}\left[l_1(\Theta(x_{l}), x_{ref})\right] = \argmax_{\Theta} \mathbb{E}\left[\log f(\Theta{(x_{l})}|x_{ref})\right],
    \label{eq:classicaleq}
\end{equation}
where $\Theta(x_{l})$ is the normal-light image generated by the model and $f$ is the probability density function conditioned on the reference image $x_{ref}$ defined as follows:
\begin{equation}
    f(x|x_{ref}) = \frac{1}{2b}\exp \left(-\frac{\left|x-x_{ref}\right|}{b} \right),\\
\label{eq:laplace}
\end{equation}

\noindent \yufei{where $b$ is a constant related to the learning rate.} However, such a training paradigm has a limitation that the pre-defined distribution (\eg, the distribution in Eq. \ref{eq:laplace}) of images \wh{is not strong enough to distinguish} between the generated realistic normally exposed image and the images with noises or artifacts such as the \wh{example} in Fig. \ref{fig:cond-prob}.

\subsection{Framework}
To this end, we propose to model the complicated distribution of normally exposed images using a normalizing flow so that the conditional PDF of a normally exposed image can be expressed as $f_{flow}(x|x_l)$. More specifically, a conditional normalizing flow $\Theta$ is used to take a low-light image itself and/or its features as input and maps a normally exposed image $x$ to a latent code $z$ which has the same dimension with $x$, \ie, $z=\Theta (x; x_l)$. By using \wh{the} change of variable theorem, we can obtain the relationship between $f_{flow}(x|x_l)$ and $f_z(z)$ \wh{as follows:}
\begin{equation}
    f_{flow}(x|x_{l}) = f_{z}\left(\Theta(x_{ref};x_l)\right) \left|\det \frac{\partial \Theta}{\partial x_{ref}} \left(x_{ref};x_l\right) \right|.
    \label{eq:f_flow}
\end{equation}

\wh{To make the model better characterize the properties of high-quality normally exposed images}, we use the maximum likelihood estimation to estimate the parameter $\Theta$. 
Specifically, we minimize the negative log-likelihood (NLL) instead of $l_1$ loss to train the model
\begin{equation}
\begin{split}
\small
    L&(x_l, x_{ref}) = -\log f_{flow}(x_{ref}|x_l) \\
        &=  -\log f_z(\Theta (x_{ref}; x_l)) 
        - \sum_{n=0}^{N-1} \log |\det 
         \frac{\partial \theta^n}{\partial z^n}(z^n; g^n(x_l))|,
\end{split}
\end{equation}
where the
\yufei{invertible network $\Theta$
is divided into a sequence of $N$ invertible layers $\{\theta^1, \theta^2,...,\theta^N\}$ 
and $h^{i+1}=\theta^i(h^i;g^i(x_l))$ is the output of layer $\theta^i$ ($i$ ranges from $0$ to $N-1$), $h^0=x_{ref}$ and $z=h^N$. $g^n(x_l)$ is the latent feature from the encoder $g$ that has the compatible shape with the layer $\theta^n$. $f_z$ is the PDF of the latent feature $z$.}

In summary, our proposed framework includes two components: an encoder $g$ which takes a low-light image $x_l$ as input and output illumination invariant color map $g(x_l)$ (which can be regarded as reflectance map inspired by Retinex theory), and an invertible network \wh{that maps} a normally exposed image to a latent code $z$. The details of the two components are introduced in the following subsections. 

\begin{figure}[tbp]
    \centering
    \subfigure[$x_l$]{\includegraphics[width=0.24\linewidth, height=0.16\linewidth]{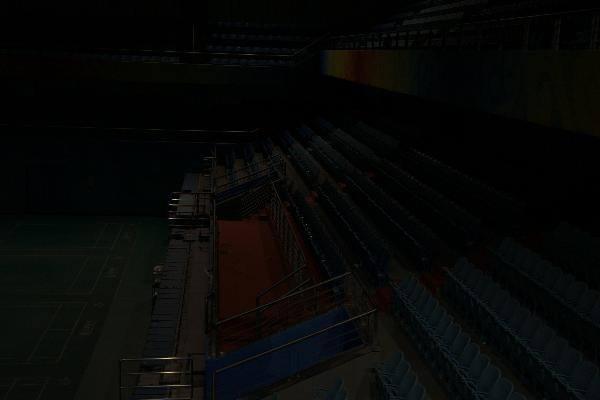}}
    \subfigure[$h(x_l)$]{\includegraphics[width=0.24\linewidth, height=0.16\linewidth]{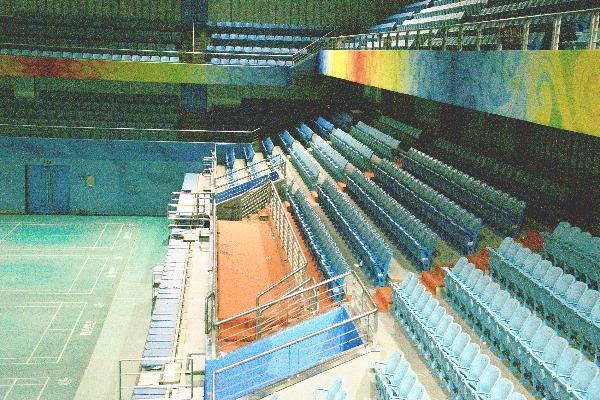}}
    \subfigure[$C(x_l)$]{\includegraphics[width=0.24\linewidth, height=0.16\linewidth]{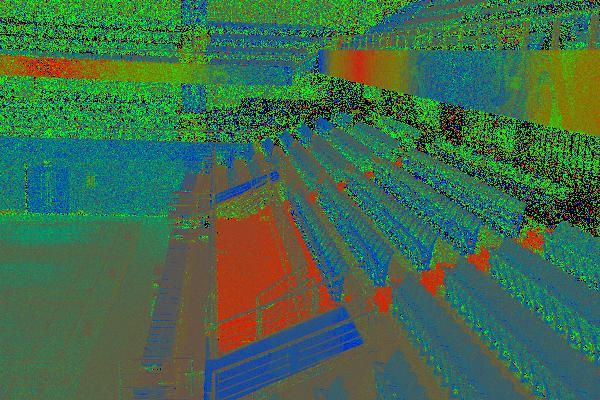}}
    \subfigure[$N(x_l)$]{\includegraphics[width=0.24\linewidth, height=0.16\linewidth]{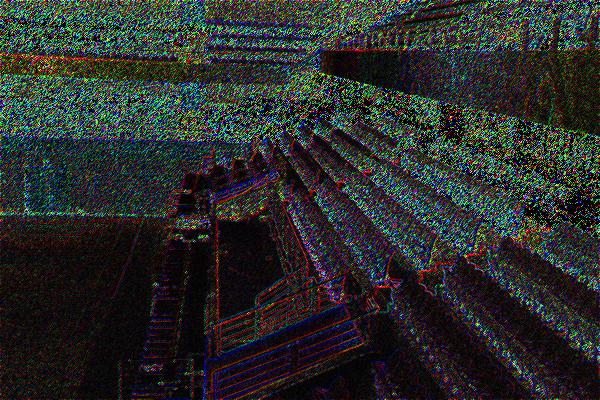}}    
    \caption{
    The components of the input for the encoder $g$. The low-light image $x_l$, low-light image after histogram equalization $h(x_l)$, color map $C(x_l)$ and noise map $N(x_l)$ are concatenated to form the input with 12 channels.
    }
    \label{fig:encoder_input}
\end{figure}

\subsubsection{Encoder for illumination invariant color map:} To generate robust and high quality illumination invariant color maps, the input images are first processed to extract useful features and the extracted features are then \wh{also} concatenated as \wh{a part of the} input of the encoder built by Residual-in-Residual Dense Blocks (RRDB)~\cite{wang2018esrgan}.
%
The detailed architecture of the encoder $g$ is in appendix due to limited space. The visualizations of each \wh{component} are shown in Fig. \ref{fig:encoder_input} and the details are as follows:

\textbf{\wh{1)} \yufei{Histogram equalized image}  $h(x_l)$:} Histogram equalization is conducted to increase the global contrast of low-light images. 
\wh{The histogram equalized image can be regarded as a more illumination invariant one.}
\wh{By including the histogram equalized image as a part of the network's input,} the network can better deal with \wh{the} areas that are too dark or bright.
\begin{figure}[htbp]
    \centering
    \subfigure[$C(x_l)$]{\includegraphics[width=0.24\linewidth, height=0.16\linewidth]{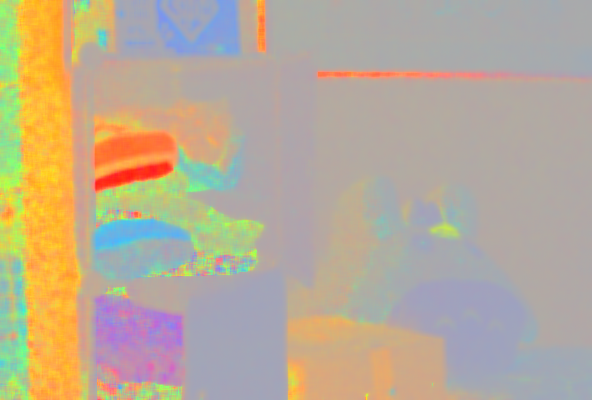}}
    \subfigure[$g(x_l)$]{\includegraphics[width=0.24\linewidth, height=0.16\linewidth]{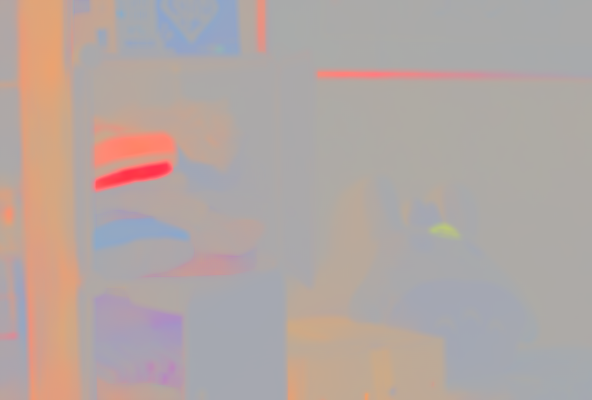}}
    \subfigure[$C(x_{ref})$]{\includegraphics[width=0.24\linewidth, height=0.16\linewidth]{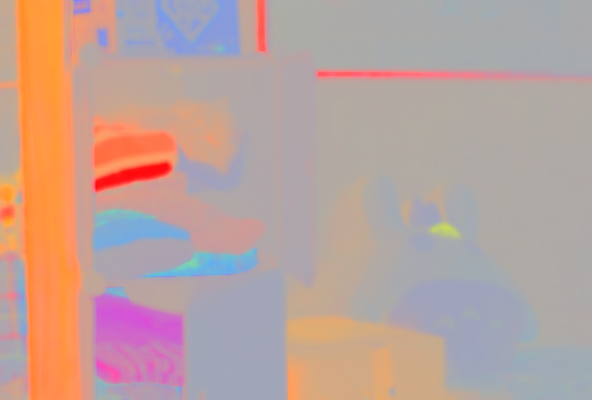}}
    \subfigure[$x_{ref}$]{\includegraphics[width=0.24\linewidth, height=0.16\linewidth]{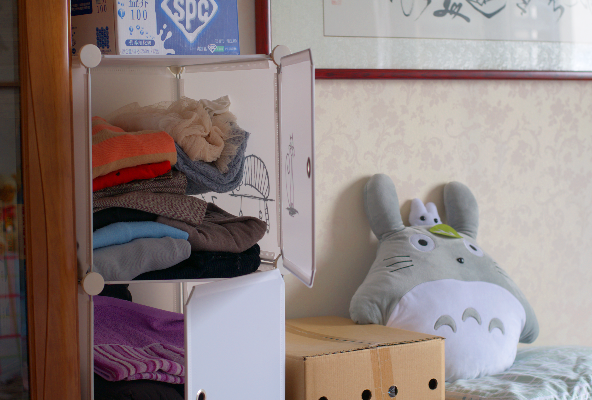}}
    \caption{
    The color map directly extracted from the low-light image $x_l$, obtain from the encoder $g$, directly extracted from the reference image $x_{ref}$, and the reference image itself.
    }
    \label{fig:colormap}
\end{figure}

\textbf{\wh{2)} Color map $C(x)$:} Inspired by Retinex theory, we propose to calculate the color map of an image $x$ as follows: 
\begin{equation}
    C(x)=\frac{x}{\text{mean}_c({x})},
\end{equation}
where $\text{mean}_c$ \wh{calculates} the mean value of each pixel among RGB channels. 
The comparison between the color map from \wh{the low-light image}, reference image, and the color map fine-tuned by the encoder $g$ is shown in Fig. \ref{fig:colormap}. 
As we can see, the color maps $C(x_l)$ and $C(x_{ref})$ are \wh{consistent to an extent} under different illumination so \wh{they} can be regarded as \wh{representations similar to the reflectance map, degraded with intensive noises in $C(x_l)$.}
We can also find that the encoder $g$ can generate a high-quality color map that \wh{suppresses} the strong noises \wh{to an extent} and \wh{preserves} the color information.

\textbf{\wh{3)} Noise map $N(x_l)$:}
To remove the noise in $C(x_l)$, a noise map $N(x_l)$ is estimated and fed into the encoder as an attention map. The noise map $N(x_l)$ is estimated as follows:
\begin{equation}
    N(x)=\max(\abs(\nabla_x C(x)), \abs(\nabla_y C(x))),
\end{equation}
where $\nabla_x$, \wh{and} $\nabla_y$ are the gradient \wh{maps} \wh{in} the directions of $x$ and $y$, \wh{where} $\max(x,y)$ is the operation that \wh{returns} the maximum value between $x$ and $y$ at the pixel channel level.

\subsubsection{Invertible network:}
Different from the encoder that learns a one-to-one mapping to extract illumination invariant color map which can be seen as \wh{the intrinsic invariant} properties of the objects, the \yufei{invertible network} aims to learn a one-to-many relationship since the illumination may be diverse for the same scenario. Our invertible network is composed of three levels, and \wh{at} each level, there are a squeeze layer and 12 flow steps. More details about the architecture can be found in the appendix.

According to our assumption that the normalizing flow aims to learn a conditional distribution of the normally exposed images conditioned on the low-light image/the illumination invariant color map, the normalizing flow should work well conditioned on both $g(x_l)$ and $C(x_{ref})$ since these two \wh{maps} are expected to be similar. To this end, we train the whole framework (both the encoder and the invertible network) in the following manner:
\begin{equation}
\begin{split}
\small
    L(x_l, x_{ref}) &=  -\log f_z(\Theta (x_{ref}; x_l)) \\
        &- \sum_{n=0}^{N-1} \log \left|\det 
         \frac{\partial \theta^n}{\partial z^n}(z^n; g^n(x_l))\right| ,
\end{split}
\end{equation}
where $f_z$ is the PDF of \wh{the} latent feature $z$ defined as follows
\begin{equation}
    f_z(z) = \frac{1}{\sqrt{2\pi}}\exp\left(\frac{-(x-r(C(x_{ref}),g(x_l)))^2}{2}\right)
\end{equation}
and \wh{$r(a,b)$} is a random selection function that is defined as follows:
\begin{equation}
r(a,b) = \left\{\begin{matrix}
a & \alpha \leq p\\ 
b & \alpha > p 
\end{matrix}\right., \alpha\sim U(0,1),
\end{equation}
in which $p$ is a hyper-parameter and we set $p$ to be 0.2 for all experiments. As shown in Fig. \ref{fig:colormap}, even without the help of pixel reconstruction loss, the encoder $g$ can learn a similar color map with the reference image.

To generate a normally exposed image using a low-light image, the low-light image is first passed through the encoder to extract the color map $g(x_l)$ and then the latent features of the encoder are used as the condition for the invertible network. 
For the sampling strategy of $z$, one can randomly select a batch of $z$ from the distribution $\mathcal{N}(g(x_L),\mathbf{1})$ to get different outputs and then calculate the mean of generated normally-exposed images to achieve better performance. 
To speed up the inference, we directly select $g(x_l)$ as the input $z$ and we empirically find that it can achieve \wh{a} good enough result. So for all the experiments, we just use the mean value $g(x_l)$ as the latent feature $z$ for the conditional normalizing flow if not specified.
\begin{figure*}[tbp]
    \centering
    \subfigure[Input]{
        \includegraphics[width=0.194\linewidth, height=0.12933\linewidth]{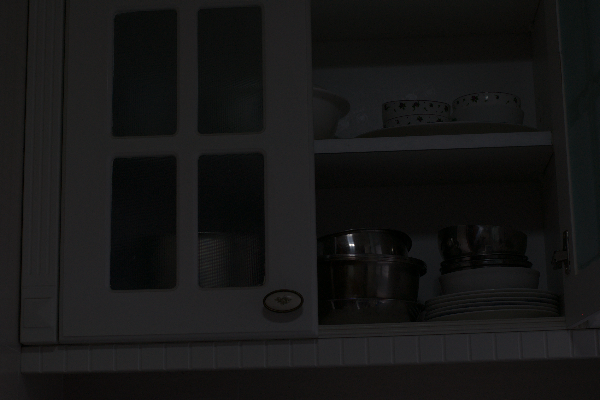}
    }\hspace{-8pt}
    \subfigure[LIME]{ 
        \includegraphics[width=0.194\linewidth, height=0.12933\linewidth]{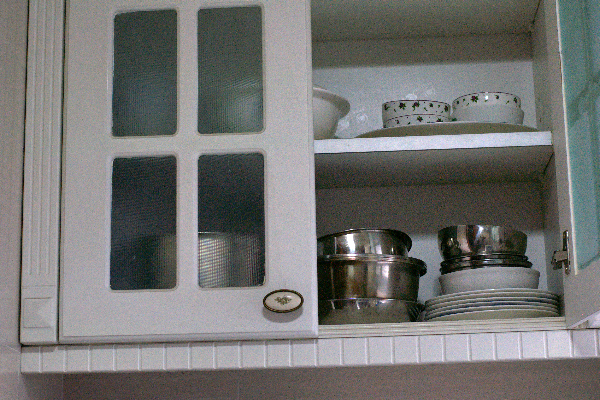}
    }\hspace{-8pt}
    \subfigure[RetinexNet]{ 
        \includegraphics[width=0.194\linewidth, ]{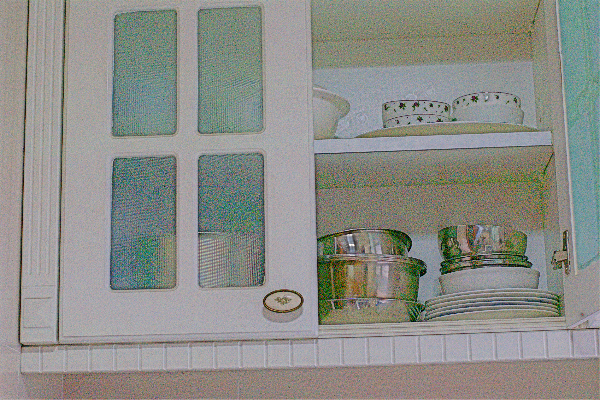}
    }\hspace{-8pt}
    \subfigure[EnlightenGAN]{ 
        \includegraphics[width=0.194\linewidth, height=0.12933\linewidth]{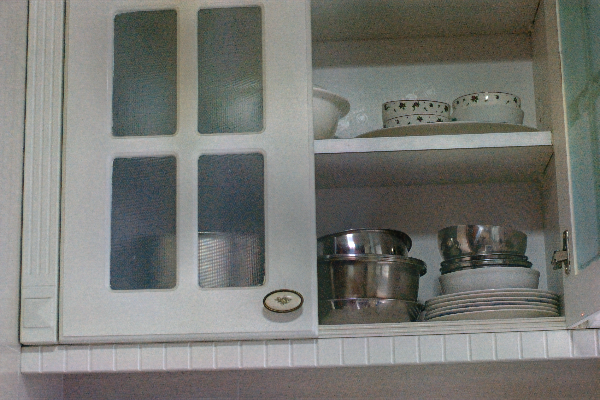}
    }\hspace{-8pt}
    \subfigure[DRBN]{ 
        \includegraphics[width=0.194\linewidth, height=0.12933\linewidth]{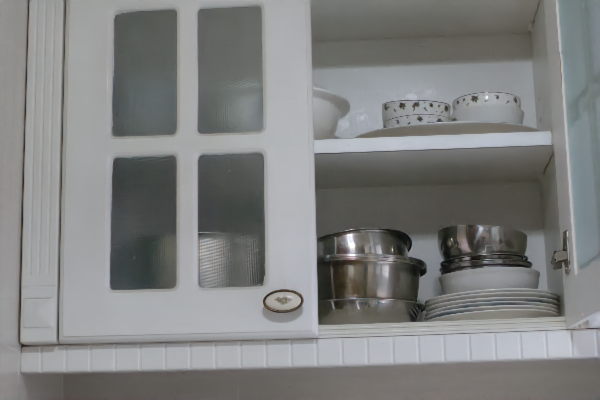}
    }
    \\
    \subfigure[Kind++]{ 
        \includegraphics[width=0.194\linewidth, height=0.12933\linewidth]{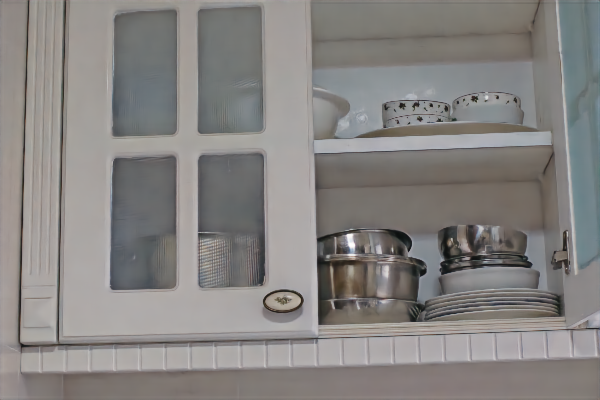}
    }\hspace{-8pt}
    \subfigure[Kind]{ 
        \includegraphics[width=0.194\linewidth, height=0.12933\linewidth]{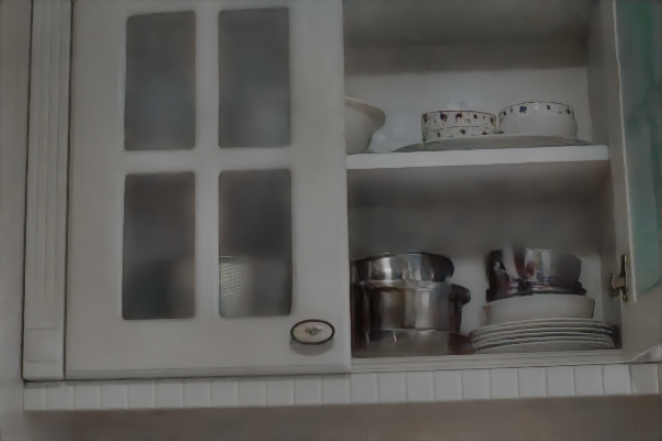}
    }\hspace{-8pt}
    \subfigure[Zero-DCE]{ 
        \includegraphics[width=0.194\linewidth, height=0.12933\linewidth]{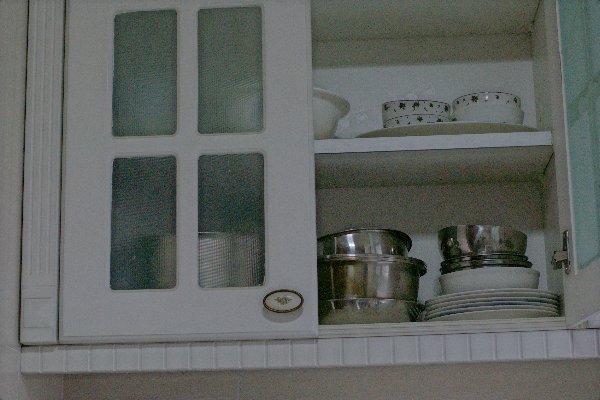}
    }\hspace{-8pt}
    \subfigure[Ours]{
        \includegraphics[width=0.194\linewidth, height=0.12933\linewidth]{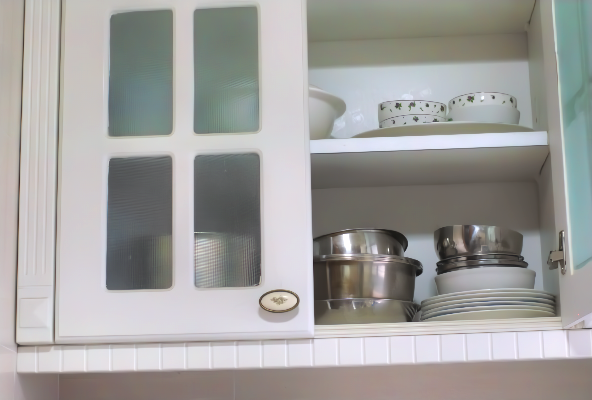}
    }\hspace{-8pt}
    \subfigure[Reference]{
        \includegraphics[width=0.194\linewidth, height=0.12933\linewidth]{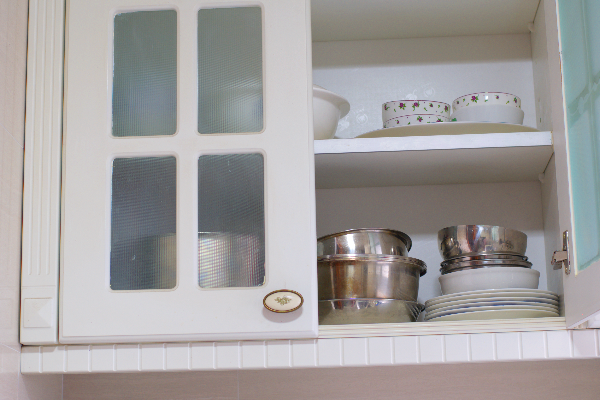}
    }
    \caption{Visual comparison with state-of-the-art low-light image enhancement methods on LOL dataset. \yufei{The normally exposed image generated by our method has less noise and artifact, and better colorfulness.}}
    \label{fig:lol-result}
    \vspace{-0.4cm}
\end{figure*}

\section{Experiments}

\wh{\subsection{Experimental Settings}}
The patch size is set to $160 \times 160$ and the batch size is set to $16$. 
We use Adam as the optimizer with a learning rate of $5\times 10^{-4}$ and without weight decay. 
\yufei{For LOL dataset}, we train the model for $3\times 10^{4}$ iterations and the learning rate is decreased with a factor of $0.5$ at $1.5\times 10^{4}$, $2.25 \times 10^{4}$, $2.7 \times 10^{4}$, $2.85 \times 10^{4}$ iterations. \
\yufei{For VE-LOL dataset, we train the model for $4\times 10^{4}$ iterations and the learning rate is decreased with a factor of $0.5$ at $2\times 10^{4}$, $3 \times 10^{4}$, $3.6 \times 10^{4}$, $3.8 \times 10^{4}$ iterations.}

\wh{\subsection{Evaluation on LOL}} 
\yufei{We first evaluate our method on the LOL datset~\cite{wei2018deep} \wh{including} 485 images for training and 15 images for testing.}
Three metrics are adopted for quantitative comparison including PSNR, SSIM \cite{wang2004image}, and LPIPS \cite{zhang2018unreasonable}. 
The numerical results among different methods are reported in Table \ref{tab:lol}. 
From \Tref{tab:lol}, we can find that our method significantly outperforms all the other competitors. 
The higher PSNR values \wh{show} that our method is capable of suppressing the artifacts and better recovering color information.
The better SSIM values demonstrate that our method better preserves the structural information with high-frequency details. 
In terms of LPIPS, a metric designed for the human perception, our method also \wh{achieves} the best performance, which indicates our method \wh{better align with} the human perception. 
\wh{The qualitative results are shown in \Fref{fig:lol-result}}
Our method achieves more promising perceptual quality by better suppressing the artifacts and \wh{revealing} image details.

\begin{table}[htbp]
    \centering
    \caption{Quantitative comparison on the LOL dataset~\cite{wei2018deep} in terms of PSNR, SSIM and LPIPS.
    \wh{$\uparrow$ ($\downarrow$) denotes that, larger (smaller) values lead to better quality.}
    }
    \scalebox{0.82}{
    \begin{tabular}{cccc}
    \toprule
      \wh{Method}  &  PSNR $\uparrow$ & SSIM $\uparrow$ & LPIPS $\downarrow$ \\
    \midrule
       Zero-DCE \cite{guo2020zero} & 14.86 & 0.54 & 0.33\\ 
       LIME \cite{guo2016lime} & 16.76 & 0.56 & 0.35 \\
       EnlightenGAN \cite{jiang2021enlightengan} & 17.48 & 0.65 & 0.32 \\
       RetinexNet \cite{wei2018deep} & 16.77 & 0.56 & 0.47 \\
       RUAS \cite{liu2021ruas} & 18.23 & 0.72 & 0.35 \\
       DRBN \cite{yang2020fidelity} & 20.13 & 0.83 & 0.16\\
       \cite{lv2021attention} & 20.24 & 0.79 & 0.14 \\
       KinD \cite{zhang2019kindling} & 20.87 & 0.80 & 0.17 \\
       KinD++ \cite{zhang2021beyond} & 21.30 & 0.82 & 0.16 \\
        \wh{LLFlow} (Ours) & \textbf{25.19} & \textbf{0.93} & \textbf{0.11} \\
    \bottomrule
    \end{tabular}}
    \label{tab:lol}
\end{table}

\subsection{Evaluation on VE-LOL}
To better evaluate the performance and \wh{generality} of our method, we further \wh{perform evaluation} on VE-LOL~\cite{liu2021benchmarking} dataset.
It is a large-scale dataset \wh{including} 2500 paired images with more diversified scenes and contents, \wh{thus is valuable for} the cross-dataset evaluation.

\begin{figure*}[tbp]
    \centering
    \subfigure[Input]{
        \includegraphics[width=0.194\linewidth, height=0.12933\linewidth]{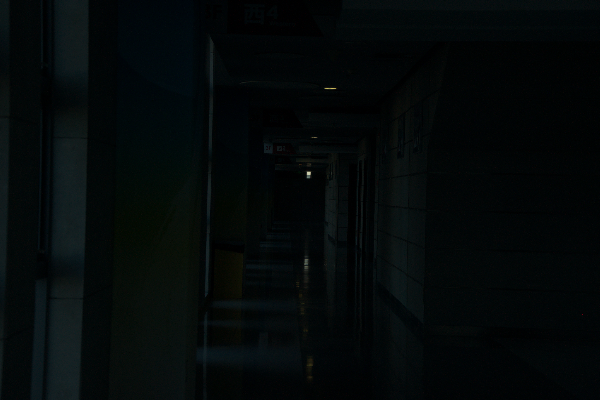}
    }\hspace{-8pt}
    \subfigure[LIME]{ 
        \includegraphics[width=0.194\linewidth, height=0.12933\linewidth]{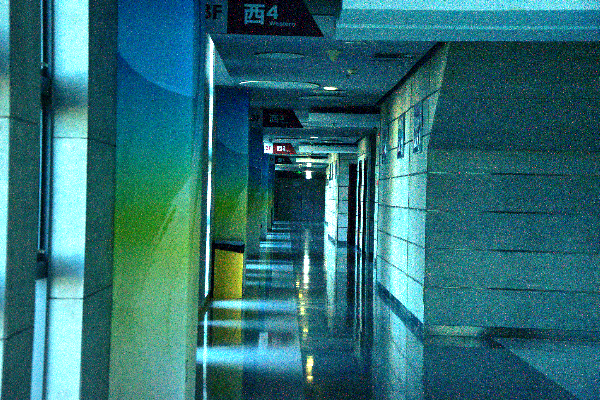}
    }\hspace{-8pt}
    \subfigure[RetinexNet]{ 
        \includegraphics[width=0.194\linewidth, ]{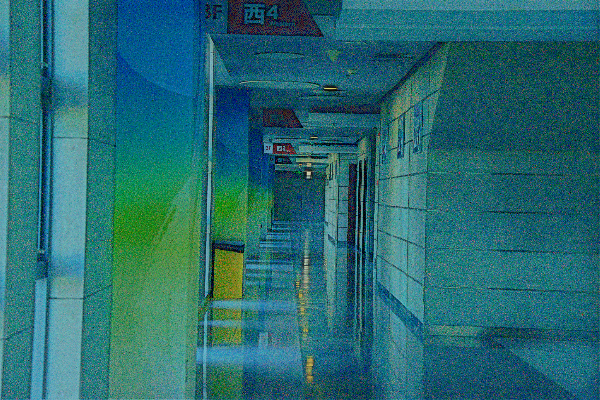}
    }\hspace{-8pt}
    \subfigure[EnlightenGAN]{ 
        \includegraphics[width=0.194\linewidth, height=0.12933\linewidth]{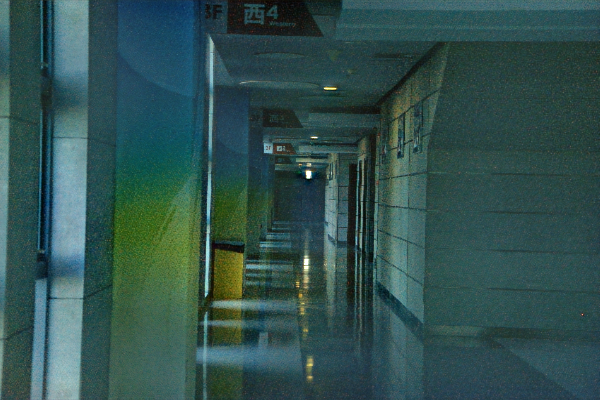}
    }\hspace{-8pt}
    \subfigure[LLNet]{ 
        \includegraphics[width=0.194\linewidth, height=0.12933\linewidth]{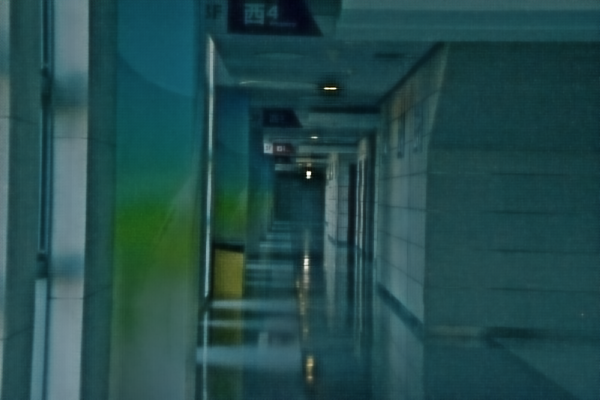}
    }
    \\
    \subfigure[Kind++]{ 
        \includegraphics[width=0.194\linewidth, height=0.12933\linewidth]{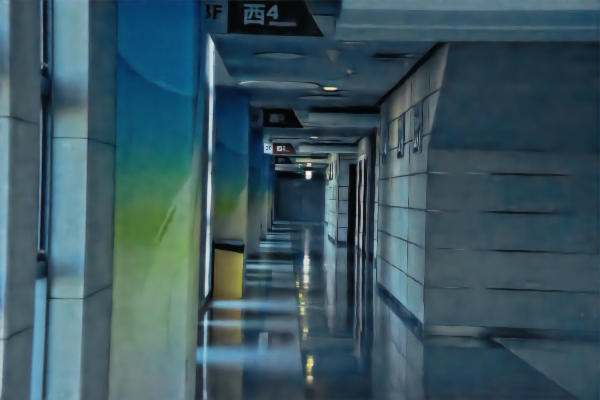}
    }\hspace{-8pt}
    \subfigure[Kind]{ 
        \includegraphics[width=0.194\linewidth, height=0.12933\linewidth]{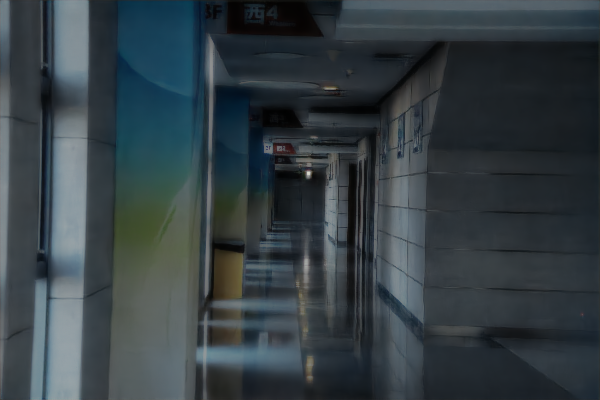}
    }\hspace{-8pt}
    \subfigure[Zero-DCE]{ 
        \includegraphics[width=0.194\linewidth, height=0.12933\linewidth]{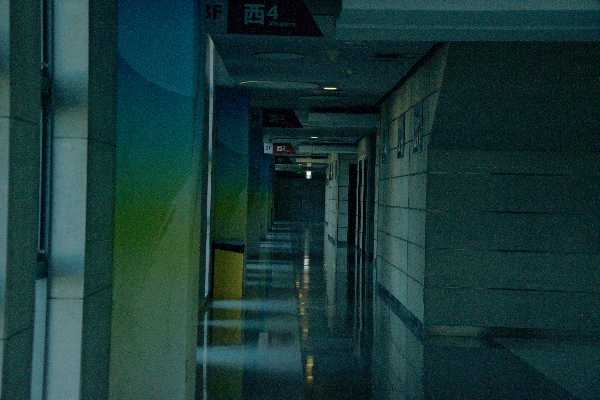}
    }\hspace{-8pt}
    \subfigure[Ours]{
        \includegraphics[width=0.194\linewidth, height=0.12933\linewidth]{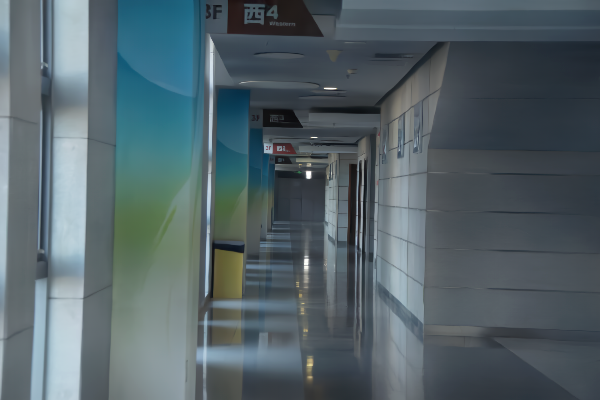}
    }\hspace{-8pt}
    \subfigure[Reference]{
        \includegraphics[width=0.194\linewidth, height=0.12933\linewidth]{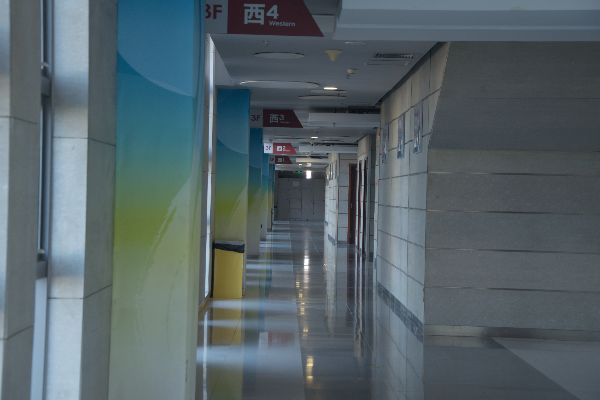}
    }
    \caption{Visual comparison with state-of-the-art low-light image enhancement methods on the real-captured set of VE-LOL dataset. }
    \label{fig:lol-v2-result-real}
\end{figure*}

\wh{\noindent \textbf{1) Cross-dataset evaluation:}}
We first evaluate the \wh{generality} of our method in a cross-dataset manner, \ie, we train our method on the LOL dataset~\cite{wei2018deep} and test the model on the testing set of VE-LOL dataset~\cite{liu2021benchmarking}.
The quantitative results are reported in Table \ref{tab:lol-v2}. From the results, our method significantly outperforms other methods in terms of all \wh{metrics}. The qualitative comparisons of real-captured image are given in \Fref{fig:lol-v2-result-real}. The results generated by our methods are with less noise and better color saturation.

\begin{table}[htbp]
    \centering
    \caption{Quantitative comparison on the VE-LOL dataset in terms of PSNR, SSIM and LPIPS. The models are trained on the training set of LOL.
    \wh{$\uparrow$ ($\downarrow$) denotes that, larger (smaller) values lead to better quality.}
    }
    \scalebox{0.82}{
    \begin{tabular}{cccc}
    \toprule
      \wh{Method}   &  PSNR $\uparrow$ & SSIM $\uparrow$ & LPIPS $\downarrow$ \\
    \midrule
    RetinexNet \cite{wei2018deep} & 14.68 & 0.5252 & 0.6423\\
    BIMEF \cite{ying2017bio} & 15.95 & 0.6386 & 0.4573 \\
    DeepUPE \cite{wang2019underexposed} & 13.19 & 0.4902 & 0.4634 \\
    JED \cite{ren2018joint} & 16.73 & 0.6817 & 0.3899 \\
    LIME \cite{guo2016lime} & 14.07 & 0.5274 & 0.4021 \\
    SICE \cite{cai2018learning} & 18.06 & 0.7094 & 0.5078 \\
    LLNet \cite{lore2017llnet} & 17.57 & 0.7388 & 0.4021 \\
    SRIE \cite{fu2016weighted} & 13.66 & 0.5509 & 0.4577\\
    KinD \cite{zhang2019kindling} & 18.42 & 0.7658 & 0.2879\\
    KinD++ \cite{zhang2021beyond} & 17.63 & 0.7994 & 0.2257\\
    Zero-DCE \cite{guo2020zero} & 21.12 & 0.7705 & 0.2480 \\
    EnlightenGAN \cite{jiang2021enlightengan} & 20.43 & 0.7921 & 0.2416 \\
    \wh{LLFlow} (Ours) & \textbf{23.85} & \textbf{0.8986} & \textbf{0.1456} \\
    \bottomrule
    \end{tabular}
    }
    \label{tab:lol-v2}
\end{table}

\begin{figure}[htbp]
    \centering
    \subfigure[]{
    \includegraphics[width=0.3\linewidth]{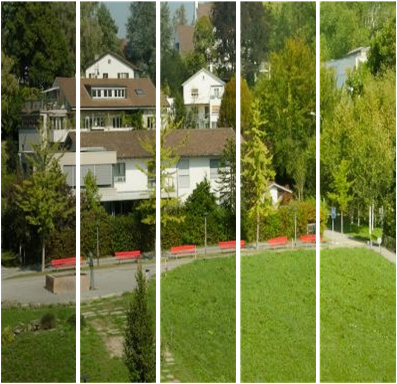}}\hspace{-1pt}
    \subfigure[]{
    \includegraphics[width=0.3\linewidth]{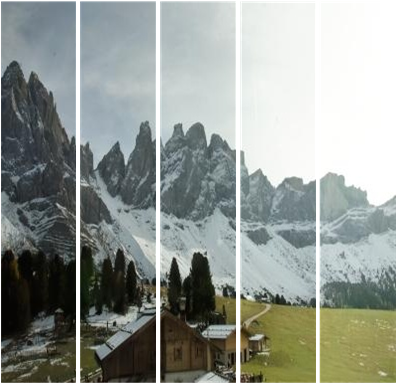}}\hspace{-1pt}
    \subfigure[]{    \includegraphics[width=0.3\linewidth]{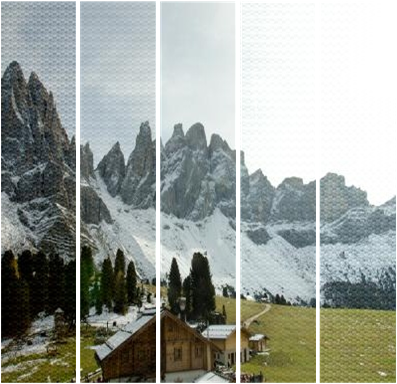}}\hspace{-1pt}
    \vspace{-0.2cm}
    \caption{(a) and (b) are the generated normally exposed images with different $z$ \wh{(monotonically, each column)} from a well-trained model. There are strong artifacts in (c) obtained from an early checkpoint of the model when it cannot well distinguish the artifacts and the variance of data. Zoom in to see details.}
    \label{fig:different_z}
\end{figure}    
\vspace{-0.4cm}



\noindent \wh{\textbf{2) Intra-dataset evaluation}}:
To further evaluate the performance of our proposed model, we compare our method with SOTA methods in an intra-dataset setting, \ie, we retrain all the methods using the training set of VE-LOL dataset and report the performance on its corresponding test set.  
The quantitative results are reported in Table \ref{tab:lol-v2-intra}. We can find that our method has the best performance and outperforms others by a large margin. Meanwhile, with the help of more diverse data, all the metrics of our method \wh{are} improved comparing with the model trained \wh{on} LOL. 
\begin{table}[htbp]
    \centering
    \caption{Quantitative comparison on the VE-LOL dataset in terms of PSNR, SSIM, and LPIPS. The models are re-trained on the training set of VE-LOL dataset.
    \wh{$\uparrow$ ($\downarrow$) denotes that, larger (smaller) values lead to better quality.}
     }
    \scalebox{0.82}{
    \begin{tabular}{cccc}
    \toprule
       \wh{Method}  &  PSNR $\uparrow$ & SSIM $\uparrow$ & LPIPS $\downarrow$ \\
    \midrule
       Zero-DCE \cite{guo2020zero} & 20.54 & 0.7786 & 0.3312 \\ 
       KinD \cite{zhang2019kindling} & 22.15 & 0.8535 & 0.2576 \\
       \wh{LLFlow} (Ours) & \textbf{26.02} & \textbf{0.9266} & \textbf{0.0996} \\
    \bottomrule
    \end{tabular}
    }
    \label{tab:lol-v2-intra}
\end{table}

\vspace{-0.4cm}

\subsection{Ablation study}
\noindent \wh{1)} \textbf{The losses estimated by our method and $l_1$:}
\yufei{
To verify our motivation that conditional normalizing flow can model a more complicated error distribution comparing with \wh{pixel-wise} reconstruction loss, we further compare the losses obtained by our method and $l_1$. As shown in~\Tref{tab:nll_table}, the image with \wh{intensive} noises and \wh{that} with \wh{slightly different} brightness have the same likelihood values under the measurement of $l_1$ loss, \wh{while the latter has much higher likelihood values under the measurement of our model than the former, which is better aligned with human perception.}
}

\begin{table}[htbp]
    \centering
    \caption{\yufei{The differences under $l_1$ and negative log likelihood (NLL) estimated by our method for images with brightness variance and strong noise. The mean values among the test set of the LOL dataset are reported in the table.}}
    \scalebox{0.84}{
    \begin{threeparttable}
    \begin{tabular}{ccc}
    \toprule
     \wh{Degradation} & NLL estimated by our method & $l_1$\\ 
    \midrule
    Reference & -6.09 & \text{N/A}\\
    Brightness reduced by 20\tnote{1}  & -5.95 & 20 \\ 
    Brightness increased by 20  & -6.15 & 20\\ 
    \yufei{Reference + random noise $r$\tnote{2}} & 4.84 & 20\\ 
    \bottomrule
    \end{tabular}
    \begin{tablenotes}
    \yufei{
    \item[1] The range of pixel value is $0-255$.
    \item[2] The random noise $r$ has the same shape with the reference image and the mean value and mean absolute value of $r$ are $0$ and $20$ respectively.}
    \end{tablenotes}
    \end{threeparttable}
    }
    \label{tab:nll_table}
\end{table}
\vspace{-0.1cm}

\noindent \wh{2)} \textbf{The effect of different $z$:} \renjie{A major advantage of our method over existing ones is that LLFlow can better encode the brightness variance into the latent space $z$. To verify the effectiveness of such strategy,} we add a constant to the extracted $g(x_l)$ from $-0.4$ to $0.4$ with a step of $0.2$. The results in \Fref{fig:different_z} demonstrate that the brightness of the image is monotonous with the value of $z$, which indicates that our model can encode the variance of the dataset, \ie, the inevitable uncertainty when collecting the data pairs.    


\begin{figure}[htbp]
    \centering
    \subfigure[]{
    \includegraphics[width=0.246\linewidth]{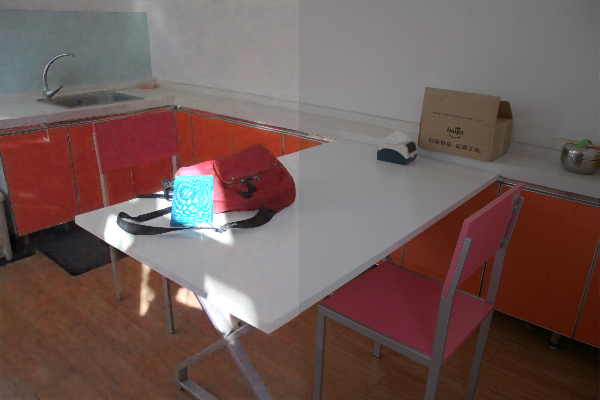}
    }\hspace{-9pt}
    \subfigure[]{
    \includegraphics[width=0.246\linewidth]{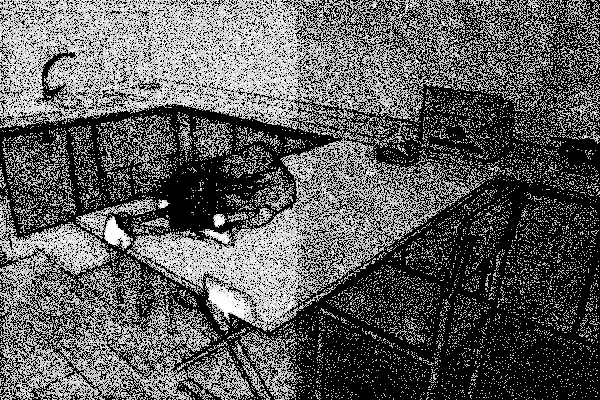}
    }\hspace{-9pt}
    \subfigure[]{
    \includegraphics[width=0.246\linewidth]{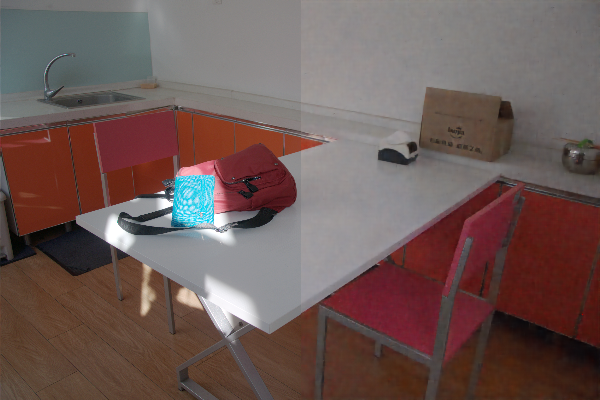}
    }\hspace{-9pt}
    \subfigure[]{
    \includegraphics[width=0.246\linewidth]{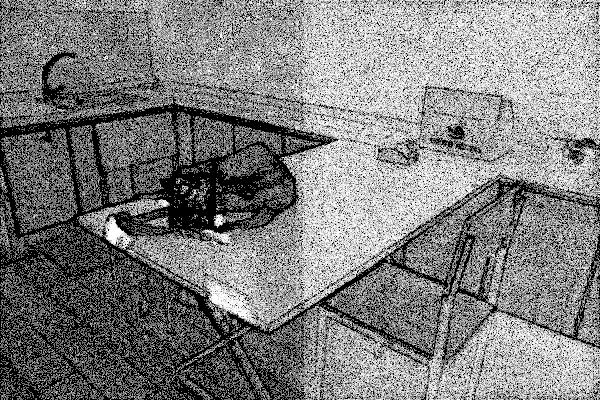}
    }\hspace{-9pt}
    \caption{
    Gradient activation map from our model. \textbf{(a)}: The stitched picture that its left half is \yufei{from not fully trained model} and its right half is from the reference image. \textbf{(b)}: The gradient activation map from (a). \textbf{(c)}: The stitched picture that its right half is \yufei{from not fully trained model} and its left half is the reference image. \textbf{(d)}: The gradient activation map from (c).}
    \label{fig:grad_act}
    \vspace{-0.6cm}
\end{figure}

\noindent \wh{3)} \textbf{The \yufei{activation} area of LLFlow:} To better understand \wh{how our model builds a more strong constraint}, we visualize the gradient activation map of our method. For a \wh{normally} exposed image $x_{high}$ which can be a reference image or the output from a low-light enhancement network from its corresponding low-light image $x_{low}$, the gradient activation map $G$ can be obtained as follows:
\begin{equation}
    G = h(||\nabla_x L(x_l,x_{high})\||_2)
\end{equation}
where $h$ is the histogram equalization operation \wh{to better visualize the results}. From the results in \wh{Fig.}~\ref{fig:grad_act}, we can find that the area with artifacts has a higher gradient activation value.
\wh{It} demonstrates that even without the reference image, our model can distinguish the unrealistic areas according to the learned conditional distribution.

\noindent \wh{4)} \textbf{The effectiveness of model components and training paradigm:} To investigate the effectiveness of our training paradigm and different components in our framework\wh{,} We evaluate the performance of our conditional encoder individually and the performance of our whole framework via training them using $l_1$ loss.

For the evaluation of our whole framework under $l_1$ \wh{loss}, we empirically find that \wh{training directly with it} cannot converge. To this end, we first pretrain the framework for \wh{1,000} iterations by minimizing the negative log likelihood $L(x_l,x_{ref})$. 
All the networks are trained \wh{with} the same batch \wh{size}, patch size, image prepossessing \wh{pipeline} \wh{in the related experiments}. 
We finetune other hyper-parameters, e.g., learning rate and weight decay, in a wide range to achieve the best performance. 

\begin{figure}[htbp]
    \centering
    \includegraphics[trim=10 35 10 500,clip, width=1\linewidth]{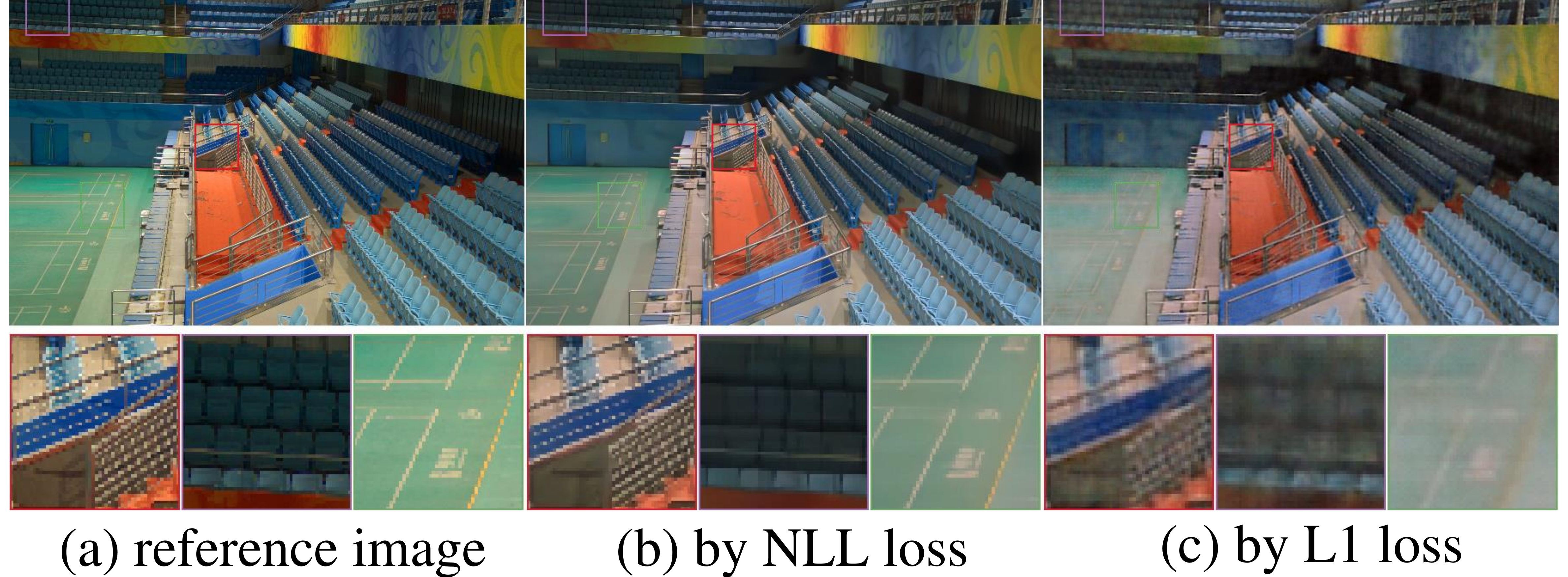}
    \caption{The effect of different training paradigm for the same network.}
    \label{fig:training-paradigm}
\end{figure}

\renjie{The results evaluated on LOL dataset~\cite{wei2018deep} are reported in Table \ref{tab:only_encoder}. The model trained by minimizing NLL loss has a huge improvement in all metrics comparing with the model trained by $l_1$ loss. A visual comparison between the results from $l_1$ loss trained model and NLL trained model are shown in Fig. \ref{fig:training-paradigm}. From the results, the model trained by $l_1$ loss produces more obvious artifacts. Both quantitative and qualitative results demonstrate the superiority of our flow-based method in modeling the distribution of images with normal brightness over a simplified pixel-wise loss.}


\begin{table}[htbp]
    \centering
    \caption{Quantitative comparison between training the model with $l_1$ and \wh{NLL} loss on the LOL dataset.
    \wh{$\uparrow$ ($\downarrow$) denotes that, larger (smaller) values lead to better quality.}
    }
    \scalebox{0.86}{
    \begin{tabular}{cccc}
    \toprule
      \wh{Loss}   &  PSNR $\uparrow$ & SSIM $\uparrow$ & LPIPS $\downarrow$ \\
    \midrule
    Only encoder (L1 loss) & 21.90 & 0.8587 & 0.1672\\
    \wh{LLFlow} (L1 loss) & 22.68 & 0.8391 & 0.2038\\
    \wh{LLFlow} (Ours) & \textbf{25.19} & \textbf{0.9252} & \textbf{0.1131} \\
    \bottomrule
    \end{tabular}
    }
    \label{tab:only_encoder}
\end{table}
\vspace{-0.3cm}

\noindent \wh{5)} \textbf{\yufei{The effect of different latent feature distributions:}}
\renjie{To evaluate the effectiveness of our proposed illumination invariant color map and different hyper-parameters $p$, we evaluate them using the LOL dataset~\cite{wei2018deep}. The results in Table \ref{tab:ablation_gamma} show that our whole model with the newly designed color map achieves better PSNR values. The higher SSIM and LPIPS values show that the color map helps improve the color and brightness consistency. }
\begin{table}[htbp]
    \centering
    \caption{The effect of different latent feature distributions. 
    \wh{$\uparrow$ ($\downarrow$) denotes that, larger (smaller) values lead to better quality.}
    }
    \scalebox{0.86}{
    \begin{tabular}{cccc}
    \toprule
        \wh{Latent Distribution} &  PSNR $\uparrow$ & SSIM $\uparrow$ & LPIPS $\downarrow$ \\
    \midrule
        \wh{LLFlow} w/o color map & 24.46 & 0.9235 & 0.1146 \\
        \wh{LLFlow} w/ color map, $p=0.5$ & 24.85 & 0.9232 & 0.1192 \\
        \wh{LLFlow} w/ color map, $p=0.2$ & \textbf{25.19} & \textbf{0.9252} & \textbf{0.1131} \\
    \bottomrule
    \end{tabular}
    }
    \label{tab:ablation_gamma}
\end{table}
\vspace{-0.4cm}

\section{Conclusion}
In this paper, we propose a novel framework for low-light image enhancement through a novel normalizing flow model. 
Compared with the existing techniques based on the pixel-wise reconstruction losses with deterministic processes, the proposed normalizing flow trained with negative log-likelihood (NLL) loss taking the low-light images/features as the condition naturally better characterizes the structural context and measures the visual distance in image manifold.
With these merits, our proposed method naturally better captures the complex conditional distribution of normally exposed images and can achieve better low-light enhancement quality, \ie, well-exposed illumination, suppressed noise and artifacts, as well as rich colors.
The experimental results on the existing benchmark datasets show that our proposed framework can achieve better quantitative and qualitative results compared with state-of-the-art techniques.

\newpage
\bibliography{ref.bib}

\end{document}